\documentstyle[seceq,supplement,epsfig]{ptptex}
\title{%               %You can use \\ for explicit line-break
Narrow rings: a scattering billiard model}
\author{%              %Use \sc for the family name
Luis {\sc Benet}\footnote{E-mail address: benet@fis.unam.mx} 
and Thomas H. {\sc Seligman}
\footnote{E-mail address: seligman@cicc.unam.mx}
}

\inst{%                %Affiliation, neglected when [addenda] or [errata]
Centro de Ciencias F\'{\i}sicas, University of M\'exico (U.N.A.M.)\\ 
62210 Cuernavaca, M\'exico\\
and\\
Centro Internacional de Ciencias A.C., 62131 Cuernavaca, M\'exico}

\recdate{%              %Editorial Office will fill in this.
\today
}

\abst{
We propose a generic mechanism for the formation of narrow rings
in rotating systems. 
For this purpose we use a system of discs rotating about a common 
center lying well outside the discs. 
A discussion of this system shows that narrow rings occur, if we 
assume non-interacting particles. 
A saddle-center bifurcation is responsible for the relevant appearance 
of elliptic regions in phase space, that will generally assume ring 
shapes in the synodic frame, which will suffer a precession in the 
sidereal frame. 
Finally we discuss possible applications of this mechanism and find that 
it may be relevant for planetary rings as well as for semi-classical 
considerations.
}

\begin{document}

\maketitle

\section{Introduction\label{sec1}}

With the discovery of ring structures around all mayor 
planets~\cite{planetaryrings} the phenomenon has past from being a 
breathtaking anomaly in the sky to a rather common feature. 
While there exists an extended body of work studying features, 
origin and destiny of planetary 
rings,~\cite{planetaryrings,brahic,ringS,dermottm,gtremaine} \  
we believe that there is room for a more general question: How 
generic is the formation of narrow rings? 
We shall here discuss a model which is quite remote from any true 
representation of planetary rings. 
The purpose will be to show that very simple rotating systems can 
and usually will have a phase space structure that allows for the 
occurrence of rings, even if the particles of such a ring do not 
move on trajectories that could be approximated by the rings shape. 
Understanding the mechanism in a simple model should reveal the 
genericity we seek.

We shall start by generalizing the model of rotating discs which 
we proposed earlier.\cite{niggi95} \  
In section~\ref{sec3} we discuss the stability of periodic orbits 
in this system and the ring structures we find as a consequence 
of stable motion for particles not interacting among themselves. 
Next, we shall see how the introduction of a second disc on a smaller 
orbit may limit the number of rings to as few as a single one.
In particular we shall see that we do not require both discs to move 
at the same angular velocity to maintain the ring structure.
We proceed finally to consider the genericity of the results obtained 
in this particular model.\cite{genrings} \  
In particular, we shall discuss the consequences of using smooth 
potentials and even attractive ones such as the $1/r$ potential 
that occurs in the planetary rings as well as possible applications.

%----------------------------------------------------------------------
\section{The rotating two-disc scattering billiard
\label{sec2}}

\begin{figure}
\noindent\centerline{
\epsfig{figure=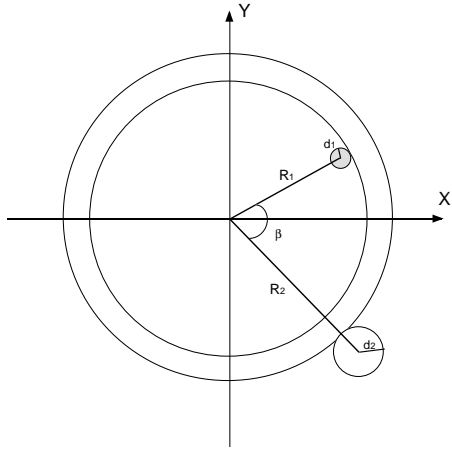,width=6cm}}
\caption{ 
\label{figuno}
Geometry of the two rotating discs scattering model: $R_i$ ($i=1,2$)
define the position of the centers of the discs and $d_i$ their radii; 
$\beta$ defines the relative positions of the discs.}
\end{figure}

We shall study the planar motion of collisionless point particles in 
an open billiard which consists of two non-overlapping discs on circular
orbits (Fig.~\ref{figuno}).
We consider discs of radii $d_i$ whose center are at distances 
$R_i$ from the origin ($i=1,2$).
For simplicity, we shall discuss the case where the discs move with the 
same angular velocity $\omega$, which permits a time independent 
Hamiltonian formulation, although the dynamics of the billiard will 
allow us to relax this condition.
The (constant) angle formed by the vectors pointing to the centers 
shall be denoted by $\beta$.
In what follows, we shall refer to the inner and the outer discs as 
disc~1 and disc~2, respectively.
The non-overlapping condition reads as $R_1+d_1<R_2-d_2$.

% Figure 1

In a rotating frame (synodic frame), the dynamics are given by the 
Hamiltonian
\begin{equation}
J={1\over 2}(p_x^2+p_y^2) - \omega (x p_y-y p_x) + V(x,y)
\label{uno}
\end{equation}
with V(x,y)=0 outside the discs, and infinite inside.
The explicit time independence of $J$ in Eq.~\ref{uno} implies that 
it is a conserved quantity of the global flow, which is known as 
the Jacobi integral.
We shall define the coordinates such that in this rotating frame the 
outer disc lies on the positive $x$-axis.
We note that, if the discs move with different angular velocities, there 
is no first integral of motion.

On a fixed frame (sidereal frame), the dynamics are 
straightforward.
A particle moves on a straight line with constant velocity until 
it hits one disc, which changes both the magnitude and the
direction of the velocity vector.\cite{niggi95} \ 
After the collision, a point particle wins (looses) energy if it 
bounces off the discs on their front (back) side.
This naive observation is precisely the key point for the construction 
of the symmetric periodic orbits of the system.
If a particle bounces radially, that is at one of the intersection 
points of the disc and the line that joins the center of rotation and 
the center of the disc, then the energy is conserved and incoming 
collision angles are equal to outgoing ones.
If we have initial conditions which display consecutive radial 
collisions with the disc, they belong to a trapped orbit.
In the synodic frame, such trapped orbits are actually periodic 
orbits which preserve the symmetry of the problem.\cite{niggi95} \ 
As shown in Ref.~\citen{benetetal99}, the fundamental periodic 
orbits in the construction of the chaotic saddle are precisely the 
symmetric periodic orbits, or deformations of them for the non-symmetric 
cases.

\begin{figure}
\noindent\centerline{
\epsfig{figure=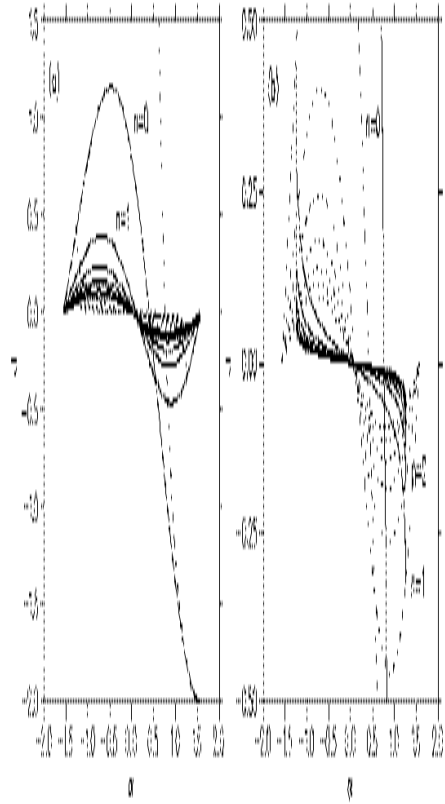,width=14cm,height=6cm,angle=90}}
\caption{ 
\label{figdos}
(a)~Periodic orbits $J_{22}$ (continuous curves) and 
$J_{12}$ for $\beta=0$ (dotted curves) for various values 
of $n$ ($R_1=1.8,\ d_1=0.1,\ R_2=3,\ d_2=1$).
Notice the divergence of $J_{12}$ as $\alpha\to 0$ for the $n=0$ case.
(b)~Enlargement of (a) showing the periodic orbits of $J_{12}$ for 
$\beta=0$ (continuous curves) for various values of $n$; some periodic 
orbits $J_{22}$ are shown as dotted curves.
The family of periodic orbits for $\beta=\pi$ are obtained from those 
for $\beta=0$ by rearranging the labels (see text).}
\end{figure}

We shall first consider either of the two discs by itself.
In terms of the Jacobi integral $J_{ii}$ ($i=1,2$), their symmetric 
periodic orbits are given by~\cite{niggi95}
\begin{equation}
J_{ii}={\omega^2 (R_i-d_i)^2\over\left[(2n+1)\pi - 2 \alpha_i\right]^2}
\left[ 2\cos^2\alpha_i-((2n+1)\pi -2\alpha_i)\sin 2 \alpha_i\right].
\label{dos}
\end{equation}
Here, $\alpha_i\in[-\pi/2,\pi/2]$ is the outgoing angle for the radial 
collision with the disc (negative values correspond to a particle 
running against the rotation), and $n=0,1,2,\dots$ denotes the number 
of full turns that the disc completes before the next collision.
These curves are plotted for some values of $n$ in Fig.~\ref{figdos}a 
as continuous lines.
Notice that $n$ defines continuous families of such periodic orbits 
with a precise hierarchical arrangement.

% Figure 2

Similar families of periodic orbits exist which alternate bounces 
among the discs.
However, symmetry considerations of these two-bounce periodic orbits 
imply that they retain the symmetric character, i.e. bounce radially, 
only for $\beta=0$ and $\beta=\pi$.
That is, except for those values of $\beta$, related periodic orbits 
are alternatively accelerated and decelerated after collisions with the
discs.
The two-bounce symmetric periodic orbits are given by
\begin{equation}
J_{12}={1\over 2}v_p^2-\omega (R_2-d_2) v_p\sin{\alpha_2}
\label{tres}
\end{equation}
where the (constant) velocity between collisions is (for $\beta=0,\pi$)
\begin{equation}
\label{cuatro}
v_p={\omega (R_2-d_2) \over \left[ n\pi+\beta +\alpha_1-\alpha_2\right]}
{\sin(\alpha_1-\alpha_2) \over \sin\alpha_1 }.
\end{equation}
In these cases $\alpha_1$ and $\alpha_2$ are geometrically related 
by the identity
\begin{equation}
\label{cinco}
{\sin\alpha_2 \over R_1+d_1}={ \sin\alpha_1 \over R_2-d_2}.
\end{equation}

Notice that Eq.~(\ref{cinco}) defines the limits 
$|\alpha_2|\le\alpha_2^{\mathrm{max}}$ for which the orbits $J_{12}$
exist, where $\sin(\alpha_2^{\mathrm{max}}) =(R_1+d_1)/(R_2-d_2)$.
The case where the equality holds correspond to tangent collisions with
disc~1.
Again, $n$ is defined as the number of turns completed by the
discs during one period of the symmetric periodic orbit.

In Fig.~\ref{figdos}b we have plotted the symmetric periodic orbits 
which involve disc~2  for $\beta=0$ for some values of $n$, as given 
directly by Eqs.~\ref{dos} and~(\ref{tres}-\ref{cuatro}).
It follows from Eqs.~(\ref{tres}-\ref{cuatro}) that for $\beta=\pi$ 
the $n$ curves correspond to the $n+1$ curves of $\beta=0$.
As it can be appreciated in Fig.~\ref{figdos}b, the curves $J_{12}$ 
display cusps at the edges of the interval where they are defined 
($|\alpha_2|=\alpha_2^{\mathrm{max}}$) and decrease monotonically.

We notice that the curves $J_{22}$ (continuous lines in 
Fig.~\ref{figdos}a) are convex or concave depending only on the sign of
$\alpha_2$, and then display local maxima and minima.
The important fact to observe here is that the derivative of
the $J_{12}$ families is strictly negative for any $n$, while 
${\rm d} J_{ii}/{\rm d}\alpha_2$ ($i=1,2$) vanishes at least 
in one point for all $n$.
This vanishing derivative is important in the context of stability of
the periodic orbits and their bifurcations.

%----------------------------------------------------------------------
\section{Stability of the $J_{22}$ symmetric periodic orbits: 
formation of rings\label{sec3}}

In order to consider the stability of the periodic orbits, we shall 
begin by analyzing those of the outer disc.
As noticed by H\'enon~\cite{henon2} for the restricted three-body 
problem, the stability properties are related to the value of 
${\rm d} J_{ii}/{\rm d} \alpha_i$.
In fact, a saddle-center bifurcation scenario associated with the 
maxima and minima of the curves $J_{ii}$ is responsible for the 
appearance or destruction of the symmetric periodic orbits 
for the rotating one-disc billiard.\cite{niggi95,dullin} \ 
Therefore, just below each of the maxima or above each of the minima, 
there exists an interval of Jacobi integrals where one of the 
corresponding periodic orbits is stable and surrounded by invariant 
tori; the other fixed point corresponds to a hyperbolic one whose 
manifolds build an incomplete horseshoe.
These maxima and minima are defined by the condition 
${\rm d} J_{22}/{\rm d} \alpha_2=0$, which can be rewritten as
\begin{equation}
\tan\alpha^*={2\over{(2n+1)\pi-2\alpha^*}}\pm 1.
\label{seis}
\end{equation}
This equation defines for a given $n$ the angles $\alpha_2=\alpha^*$
where the maxima or minima are found, independently of the
specific geometrical parameters considered in the model.

% Figure 3
\begin{figure}
\noindent\centerline{
\epsfig{figure=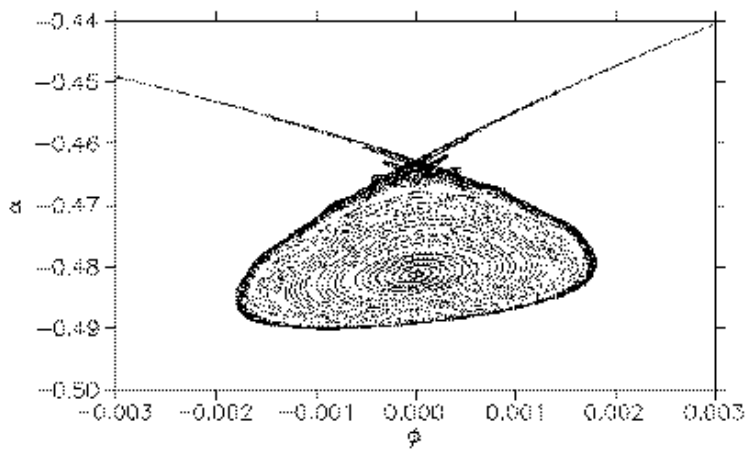,height=7cm,angle=0}}
\caption{ 
\label{figtres}
Poincar\'e surface of section showing a stability island ($J=1.173$).
An ensemble of initial conditions started within this stable region
will not escape along scattering trajectories.}
\end{figure}

In Fig.~\ref{figtres} we present the surface of section for the one-disc 
rotating scattering billiard for $J_{22}=1.173$, 
illustrating the regions of stable motion just described.
We have chosen the parameters $R_2=3$ and $d_2=1$ ($\omega=1$).
The surface of section is defined here by the angles $\alpha$ and 
$\phi$.
The latter angle characterizes the position of the collision point on 
the disc.\cite{niggi95} \ 
There are only two fixed points of period one for this value of the 
Jacobi integral, which correspond to the periodic orbits near the
maximum of the $n=0$ case (see Fig.~\ref{figdos}a).
One of these fixed points is stable and appears surrounded by KAM tori.
These KAM tori as well as some regions of chaotic motion are bounded 
dynamically by the invariant manifolds of the partner 
hyperbolic fixed point.
This implies that, even-though the problem is actually a scattering 
billiard system with a (hard disc) repulsive potential, there exist 
initial conditions of stable motion which are trapped by successive 
collisions with the rotating disc.

The KAM tori shown in Fig.~\ref{figtres} are robust and persist 
under small changes of the Jacobi integral. 
For larger changes, bifurcations occur but those of the central 
elliptic point are at first hidden from scattering trajectories 
by surviving KAM surfaces.
This scenario persists under variations of the Jacobi integral until the 
formal parameter describing the (local) binary horseshoes development 
attains the value $2/3$ (see Refs.~\citen{rueckjung,junglippsel} for 
details), which indeed defines a continuous interval for $J$ 
where bounded trajectories exist. 
As we further change the value of the Jacobi integral we reach stages 
of the incomplete horseshoe that correspond to hyperbolic structures 
and others where very small stable islands reappear.

In the present context, we shall concentrate on a set of initial 
conditions that remain trapped under small perturbations.
That is, we consider all those initial conditions which belong to 
a bounded region, without considering if their actual motion is 
periodic, quasi-periodic or chaotic.
This set of initial condition has strictly positive measure, i.e. 
there is a non-zero probability for generic initial conditions of 
being within this set~\cite{measure}.
Note that this situation is essentially different from the hyperbolic 
one, where at most a Cantor set of measure zero is trapped whether 
the horseshoe is complete or incomplete~\cite{rueckjung}.

We shall stress that, for a generic ensemble of initial conditions 
defined on the whole phase space, which is not restricted to a fixed 
Jacobi integral shell, most initial conditions will escape to infinity.
However, particles whose initial conditions are chosen close enough to 
the stable periodic orbits and within the bounded regions described 
above, will be trapped by collisions with the disc.
This observation is fundamental for understanding the occurrence of 
patterns or {\it rings} formed by infinitely many particles whose 
initial  conditions are started precisely on dynamically bounded 
regions found in distinct intervals of the Jacobi integral.

% Figure 4
\begin{figure}
\noindent\centerline{
\epsfig{figure=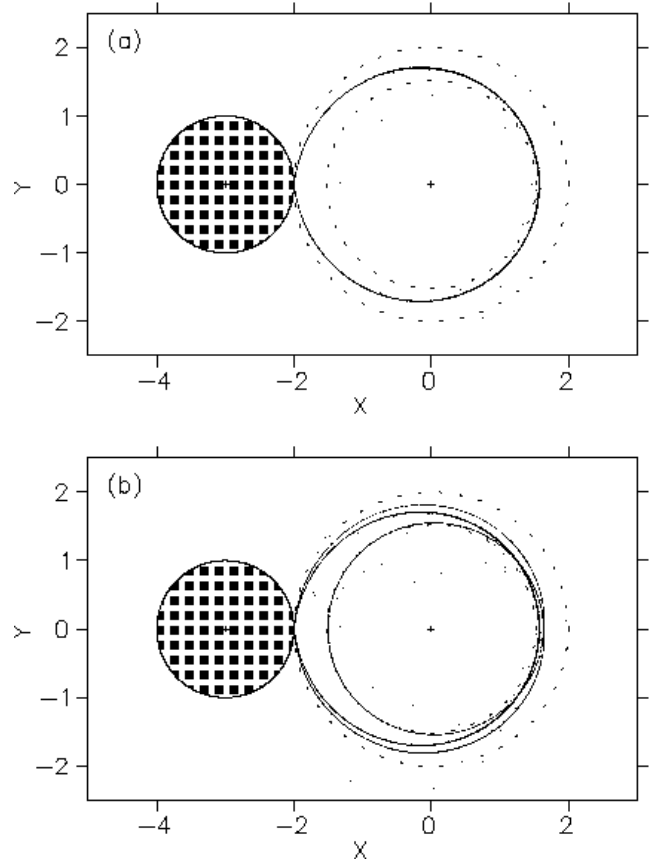,width=11.4cm,angle=90}}
\caption{ 
\label{figcuatro}
Examples of rings in the sidereal frame associated with the outer disc 
(shaded regions) for the rotating two-disc scattering problem. 
The parameters of the inner disc are defined to filter all but the 
rings shown (the parameters of the outer disc are the same as in Fig.~2).
(a)~Ring corresponding to the $n=1$ minimum of the $J_{22}$ curves
($R_1=1$, $d_1=0.52$); (b)~corresponding rings for the $n=1,2$ minima 
($R_1=0.96$, $d_1=0.52$).
The dashed circles correspond to the inner and outer circles defined by
the orbits of the outer and inner discs, respectively.}
\end{figure}

In Fig.~\ref{figcuatro}a we present in the sidereal frame the ring 
formed by an ensemble of initial conditions whose Jacobi integral 
belongs to a neighborhood of the minima of $J_{22}$ for $n=1$. 
Figure~\ref{figcuatro}b displays the set of rings formed by the 
minima of $J_{22}$ for $n=1,2$.
In these figures, each point represents at a given time the position 
of a single particle which moves on a rectilinear trajectory.
These particles have collisions with the disc at different times 
and in any point along the disc's circular orbit (sidereal frame).
The loop shown in Figs.~\ref{figcuatro}b is associated with the 
$n=2$ ring.
It results from particles on different rectilinear trajectories (with 
approximately the same Jacobi integral) which come closer during some 
interval of time.
In general, there will be $n$ loops associated with the retrograde
($\alpha_2<0$) $J_{22}$ symmetric periodic orbits, while there will 
be $n-1$ loops for the corresponding prograde orbits ($\alpha_2>0$).
The isolated points in both figures correspond to initial conditions,
that were not located in the bounded regions, but have not escaped after 
the time at which the computation was cut. 
Another remarkable aspect of Figs.~\ref{figcuatro} is that instead of 
viewing it as a stroboscopic map of many non-interacting particles in 
the sidereal frame, we could think of the narrow bands of points as 
curves, and would obtain a picture of the elliptic central orbits of 
the stable islands of a single particle in the synodic frame.

We can ask about possible structure in the rings of our model.
First of all, different rings may coexist, as we see in 
Fig.~\ref{figcuatro}b.
Their particles move at different speeds, and thus probably are not 
good candidates for the braids of planetary rings.
However, we could expect them to be important, say in a semi-classical 
context, where we might find interference between two structures. 
Another type of structure is related to the period-doubling scenario, 
that happens as the elliptic orbit bifurcates. 
There, several strands of particles may be intertwined, while they
move roughly at the same speed and maybe confined to a single ring
by surviving exterior KAM surfaces. 
The details of the structures found may be quite complicated and will 
depend on the dynamics. 
Yet the very fact that such structures generically appear is significant.

%----------------------------------------------------------------------
\section{Stability of the rings for the two-disc model\label{sec4}}

\begin{figure}
\noindent\centerline{
\epsfig{figure=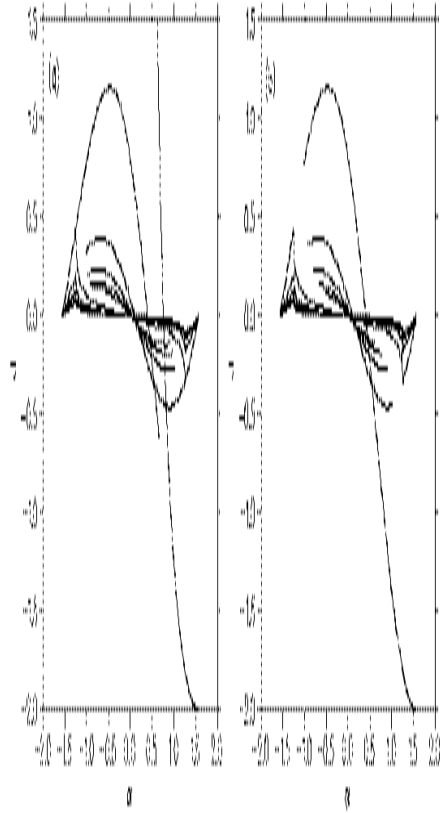,width=14cm,height=6cm,angle=90}}
\caption{ 
\label{figcinco}
Symmetric periodic orbits for the two-disc model including
the finite size effects of the inner disc. (a)~$\beta=0$; 
(b)~$\beta=\pi$.}
\end{figure}

It is important to stress that the appearance of rings as described 
above requires one rotating disc only.
Indeed infinitely many rings exist with ever increasing numbers of loops,
which give rise to arbitrarily complicated structures in wide rings.
For the full two disc rotating billiard such rings may appear associated 
with either one of the discs.
In what follows we shall not consider the rings associated with the 
inner disc only, as we are interested in narrow rings between the two 
discs that are actually influenced by both of them.
We shall thus concentrate on the periodic orbits associated with the 
outer disc, which may or may not have collisions with the inner one.
The characteristic curves for the $J_{22}$ and $J_{12}$ symmetric 
periodic orbits are shown in Fig.~\ref{figcinco} for angles 
$\beta=0,\pi$ between the discs; the size effects of the inner disc 
are taken into account ($R_1=1.8$, $d_1=0.1$).

% Figure 5

As mentioned in Section~\ref{sec2}, the $J_{12}$ (Fig.~\ref{figdos}a) 
characteristic curves for families that bounce on both discs decrease 
monotonically as a function of $\alpha_2$.
In fact, it can be shown that ${\rm d} J_{12}/{\rm d}\alpha_2$ is 
strictly non-zero.
Therefore, these families of periodic orbits do not undergo any 
bifurcation as a function of the Jacobi integral. 
Yet it was near the bifurcations that we found stable islands for 
the one-disc orbits.
Indeed we do not find any for the two-disc orbits $J_{12}$ and, as 
we shall see in the next section, there is no reason to expect any.
By consequence, these families of periodic orbits do not contribute
to the formation of rings. 
Nevertheless, their presence is intimately related to the disappearance 
of orbits of the $J_{22}$ type, as can readily be inferred from 
Fig.~\ref{figcinco}.
This erosion due to the inner disc will in general eliminate some of 
the extremal points of the characteristic curves; if a sufficiently 
large interval near the extrema is cut the stable island will no 
longer exist. 
To see this, consider initial conditions belonging to some symmetric 
periodic orbits $J_{22}$.
Depending on the geometrical parameters of inner disc, some of these 
orbits may collide with the inner disc.
The cusps displayed by the $J_{12}$ families indicate the tangent 
collisions with the inner disc (see Figs.~\ref{figdos} and~\ref{figcinco}).
The particles that collide with the inner disc will loose the needed 
correlations to build a symmetric periodic orbit, so after a few more 
collisions these particles will typically escape. 

\begin{figure}
\noindent\centerline{
\epsfig{figure=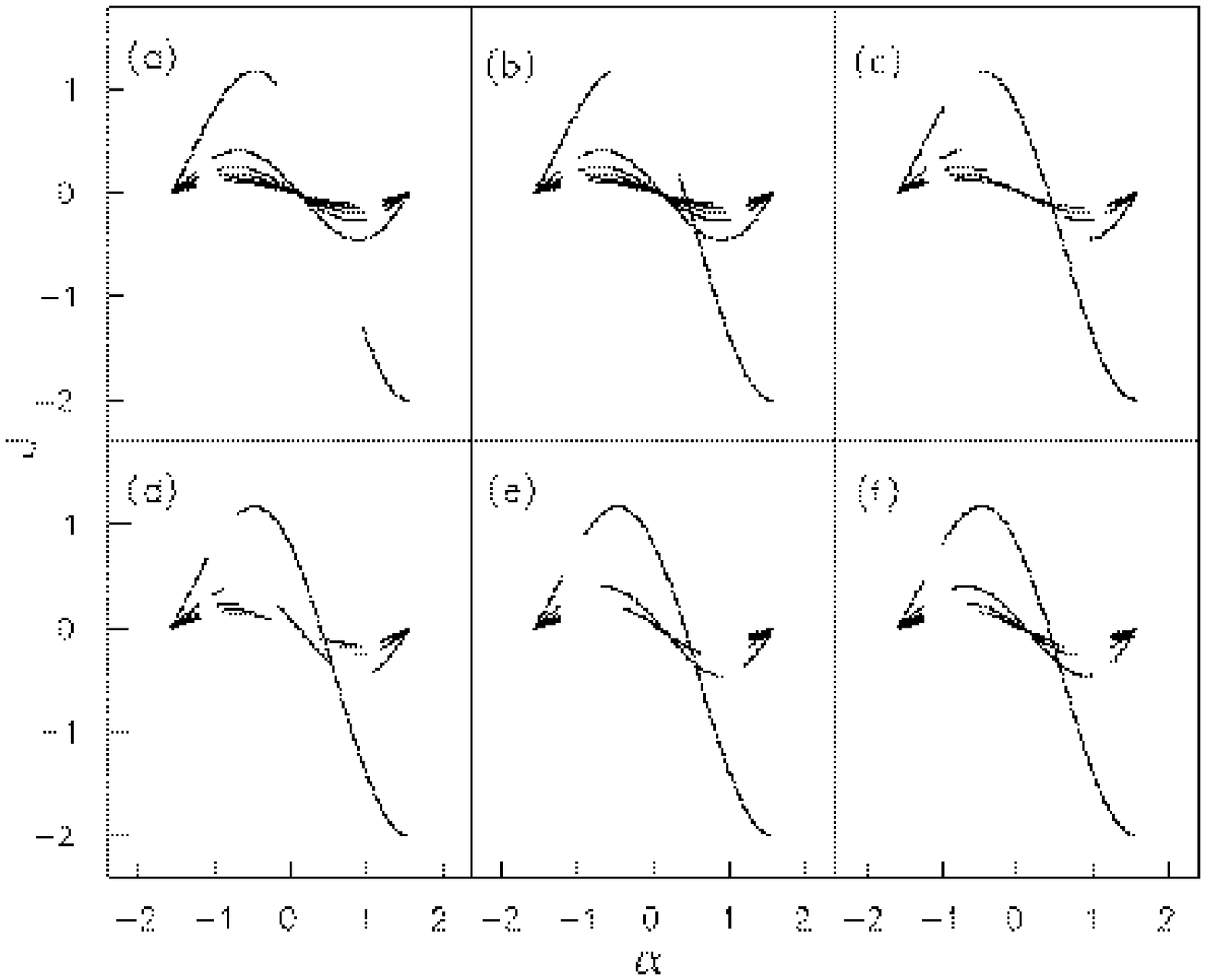,width=14.5cm,angle=90}}
\caption{
\label{figseis}
Periodic orbits of the $J_{22}$ family ($n=0,1...4$) for different
values of $\beta$. (a)~$\beta=5^\circ$,  (b)~$\beta=10^\circ$,  
(c)~$\beta=30^\circ$, (d)~$\beta=45^\circ$, (e)~$\beta=90^\circ$ 
and (f)~$\beta=135^\circ$.}
\end{figure}

A gap in the characteristic curves for the $J_{22}$ orbits thus occurs.
Its location will depend on the angle and radial distance at which the 
two discs are positioned, while its size will depend on the radius of 
the inner disc.
To illustrate this more clearly Fig.~\ref{figseis} shows the $J_{22}$ 
curves only, for some angles $\beta$.

% Figure 6

If the inner disc moves at a speed that is incommensurable with 
the outer one we find a vastly simplified picture: It will sweep all 
periodic orbits that penetrate the circle described by the outer edge 
of this disc. 
This erosion mechanism is governed in our model by 
$\alpha_2^{\mathrm{max}}$ compared to the solutions of Eq.~\ref{seis}
(which are independent of the geometrical parameters of the model).
Its typical time scale depends on the size of the inner disc and 
the time dependence of $\beta$.
The absolute value of the solutions of Eq.~(\ref{seis}) converges 
to $\pi/4$ as $n\to\infty$, from above for prograde and from below 
for the retrograde orbits.
This arrangement of the solutions implies that the lowest prograde 
solutions will be the first to survive, if we start with very small
radial distance of the discs and gradually increase it.~\cite{genrings}\ 
Note though that the $n=0$ curve in Figs.~\ref{figdos} has no stable 
prograde periodic orbits in this model (it corresponds to a whispering 
gallery orbit). 
If we want to restrict our considerations to a few rings we are 
forced to look at the prograde part of the branches for the lowest 
$n$ starting from $n=1$.
Indeed, in Fig.~\ref{figcuatro}a we see a single prograde ring with 
a circle drawn exactly in the path of the outer edge of the inner 
disc that ejects all particles that might be associated with other 
values of $n$ or with retrograde motion. 
In Fig.~\ref{figcuatro}b the prograde rings for $n=1$ and $n=2$ survive. 
These figures where generated by the dynamics with two discs 
present and somewhat scattered initial conditions.
The points that are not on the rings had unstable initial conditions 
and are on outgoing scattering orbits, but have not yet left the domain 
of the figure for the time we integrated the equations of motion.

Note that for different frequencies of the two discs the Jacobi 
integral is no longer an invariant of the entire system, but in 
the subspace of phase space where our rings live the integral 
is still conserved.

%-----------------------------------------------------------------------
\section{Genericity of the scheme and its importance for planetary rings and
other applications\label{sec5}}

If we consider Fig.~\ref{figdos}a, we see that symmetric periodic 
orbits of the $J_{22}$ family only exist between an upper and a 
lower limit for the Jacobi integral $J$.
The stable islands discussed in the previous sections are a consequence 
of the generic scenario of saddle-center bifurcations.
Thus we expect and indeed find at each branch point an elliptic an a
hyperbolic periodic orbit. 
The elliptic one will then turn into an inverse hyperbolic fixed point
via the period doubling route.

How generic is this scenario?
The presence of the cusp where two-disc orbits and one disc orbits 
touch in their characteristic curves is certainly non-generic. 
Thus the instability of the two-disc orbits and the behaviour near 
that cusp will have to be discussed in each particular case with care.
As long as the cusp is not too near the maximum of the generic curves,
the saddle-center bifurcation will proceed generically.
Cases where the two are close will need special attention.

If we consider potential mountains instead of hard discs, we will find 
an alternative generic scenario; the sudden bifurcation\cite{tgo93} 
associated with the top of the mountains. 
Yet, the scenario we discuss will still exist for trajectories that 
do never reach the region of phase space where the sudden bifurcation 
occurs.

The question becomes more involved if we consider attractive potentials. 
In view of the origin of this problem in planetary rings we shall focus 
on the $1/r$ case. 
We know from H\'enon's work~\cite{henonmu0} and some of our 
own~\cite{celmec2} that the so called consecutive collision orbits govern 
the entire chaotic saddle for the restricted three-body problem with small 
mass parameter $\mu$, i.e. small ratio between the masses of the lighter 
and heavier binary components.
These collision orbits are a good approximation both for the $1/r$ problem 
with vanishing mass parameter and for the hard disc problem with vanishing 
disc radius. 
In fact, they coincide for vanishing disc radius, when we consider an 
overall central $1/r$ attractive potential.
Therefore, the two problems are not quite as distinct as one might believe.

\begin{figure}
\noindent\centerline{
\epsfig{figure=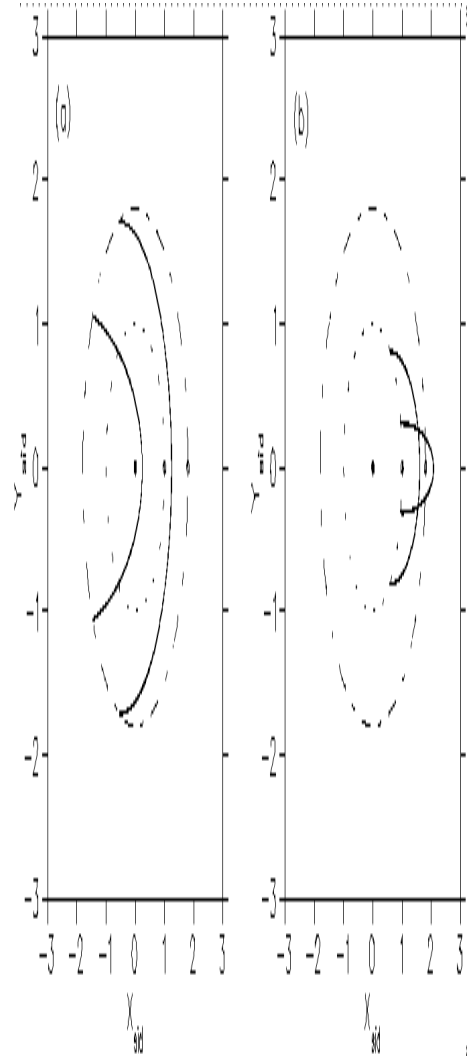,width=14cm,height=6cm,angle=90}}
\caption{
\label{figsiete}
Examples of consecutive collision orbits in the sidereal frame 
generated by collisions with (a)~the outer moonlet and, (b)~the
inner one.
The dotted and dashed circles represent the circular trajectories 
of the moonlets.
Note that in each picture one of the orbits will be swept by the
other moonlet.}
\end{figure}

The main difference is that these so called collision orbits may extend 
both outward and inward from a moonlet. 
If we consider the orbits that extend inward from the outer moonlet, 
we are in a situation quite analogous to the one we found for the discs, 
and the inner moonlet will sweep the ones that reach inside its orbit. 
If this last observation is to be effective in finite time, we have 
to pass to finite but small $\mu$. 
For the orbits that reach outward from a moonlet, the two will 
interchange their role. 
The outer one will sweep the collision orbits of the inner moonlet 
that reach outward beyond the orbit of the outer one.
In Fig.~\ref{figsiete} we illustrate, in the sidereal frame, both 
situations with one collision orbit that intersects the orbit of the 
other moonlet and another that does not. 
Clearly the zero mass approximation gives no information about stability. 
Calculations to shed light on this aspect are under way.

% Figure 7

Thus, despite of the caveats we pronounced initially, the mechanism we
discovered may well be of relevance for the structure of planetary rings.
In particular due to the generic nature of the elliptic regions, these 
are expected to be more stable against perturbations than the orbits 
of a super integrable system, and could thus be associated to a 
prediction that older ring structures would tend to consist of narrow rings.

Returning to the more general aspects, the genericity of the occurrence,
which we suspect also for attractive potentials, could imply importance 
in other systems such as rotating molecules or high spin states in 
deformed nuclei in the framework of semi-classical theory. 
In this case, the mean field potential can be assumed to rotate 
stiffly and the survival of orbits may depend on the detailed geometry 
we face. 
To simplify matters in this situation, the Jacobi integral is a constant 
of the motion.
This may be of interest when the adiabatic or Born-Oppenheimer 
approximation breaks down. 
In nuclei this can well occur for high spin states; the deviations of 
level statistics from the compound nucleus prediction observed for
such systems~\cite{experimento} was attributed to a harmonic 
component simulated by the centrifugal force, but it is worthwhile to 
consider the effect of integrable islands of the type mentioned. 
In molecules it seems less obvious to find a place for non-adiabatic 
effects, yet Rydberg molecules or molecules with exotic leptons 
may provide interesting examples.

%----------------------------------------------------------------------
\section*{Acknowledgments}
The authors want to express sincerely their gratitude for the useful
discussions and critical remarks to F. Bonasso, C. Jung and F. Leyvraz.
This work was partially supported by the DGAPA (UNAM) project IN-102597
and the CONACYT grant 25192-E.

%----------------------------------------------------------------------


\begin{thebibliography}{99}

%%%%%%%%%%%%%%%%%%%%%%%%%%%%%%%%%%%%%%%%%%%%%%%%%%%%%%%%%%%%%
% Some macros are available for the bibliography:
%   o for general use
%      \JL : general journals          \andvol : Vol (Year) Page
%   o for individual journal 
%      \PR  : Phys. Rev.               \PRL : Phys. Rev. Lett.
%      \NP  : Nucl. Phys.              \PL  : Phys. Lett.
%      \JMP : J. Math. Phys.           \CMP : Commun. Math. Phys.
%      \PTP : Prog. Theor. Phys.       \JPSJ: J. Phys. Soc. Jpn.
%      \JP  : J. of Phys.              \NC  : Nouvo Cim.
%      \IJMP: Int. J. Mod. Phys.       \ANN : Ann. of Phys.
% Usage:
%   \PR{D45,1990,345}            ==> Phys.~Rev.\ {\bf D45} (1990), 345
%   \JL{Phys.~Lett.,A30,1981,56} ==> Phys.~Lett.\ {\bf A30} (1981), 56
%   \andvol{B123,1995,1020}      ==> {\bf B123} (1995), 1020
%%%%%%%%%%%%%%%%%%%%%%%%%%%%%%%%%%%%%%%%%%%%%%%%%%%%%%%%%%%%%


\bibitem{planetaryrings} R. Greenberg and A. Brahic (eds.), 
{\it Planetary rings}. The University of Arizona Press, Tucson (1984).

\bibitem{brahic} A. Brahic, in {\it Formation of Planetary Systems} 
(A. Brahic, ed.), Centre National D'Etudes Spatiales, 
Cepadues-Editions, Toulousse (1982).

\bibitem{ringS} B.A. Smith, and the Voyager imaging team,
\JL{Science,212,1981,163}; \andvol{215,1982,504}.

\bibitem{dermottm} S.F. Dermott, C.D. Murray and A.T. Sinclair,
\JL{Nature,284,1980,309}.\\
S.F. Dermott, \JL{Nature,290,1981,454}.

\bibitem{gtremaine} P. Goldreich and S. Tremaine, \JL{Nature,277,1979,97}.

\bibitem{niggi95} N. Meyer {\it et al.}, J. Phys A: Math. Gen. 
{\bf 28} (1995), 2529.

\bibitem{genrings} L. Benet and T.H. Seligman, to be published.

\bibitem{benetetal99} L. Benet, C. Jung, T. Papenbrock and T.H. 
Seligman, \JL{Physica D,131,1999,254}.

\bibitem{henon2} M. H\'enon, \JL{Ann. Astrophys.,28,1965,992}.

\bibitem{dullin} H.R. Dullin, \JL{Nonlinearity,11,1998,151}.

\bibitem{rueckjung} B. Rueckerl and C. Jung, 
J. Phys A: Math. Gen. {\bf 27} (1994), 55.

\bibitem{junglippsel} C. Lipp, C. Jung and T.H. Seligman, 
in {\it Proc. of the Fourth Int. Wigner Symp.}, 
N.M. Atakashiev, T.H. Seligman and K.B. Wolf (eds.) World Scientific, 
Singapore (1996).

\bibitem{measure} A.J. Lichtenberg and M.A. Lieberman, {\it Regular 
and Stochastic Motion}, Applied Mathematical Sciences {\bf 38},
Springer-Verlag, New York (1983).

\bibitem{tgo93} T. T\'el, C. Grebogi and E. Ott, \JL{Chaos,3,1993,495}.

\bibitem{henonmu0} 
M. H\'enon, Bull. Astronom. (serie 3) {\bf 3} (1968), 377.\\
M. H\'enon, {\it Generating Families in the 
Restricted Three-Body Problem}, Lecture Notes in Physics {\bf m52}, 
Springer-Verlag, New York (1997).

\bibitem{celmec2} 
L. Benet, T.H. Seligman and D. Trautmann,
\JL{Celest. Mech. Dynam. Astron.,73,1999,167}.

\bibitem{burns} R.A. Kolvoord and J.A.Burns, \JL{Icarus,95,1992,253}; 
\andvol{99,1992,436}.

\bibitem{experimento} T. v.Egidy, A.N. Behkamy and H.H. Schmidt,
\JL{Nucl. Phys. A,454,1986,109}.

\end{thebibliography}
\end{document}